\documentclass[sigconf]{acmart}

\usepackage{multirow}
\usepackage{balance}

\copyrightyear{2023}
\acmYear{2023}
\setcopyright{rightsretained}
\acmConference[RecSys '23]{Seventeenth ACM Conference on Recommender Systems}{September 18--22, 2023}{Singapore, Singapore}
\acmBooktitle{Seventeenth ACM Conference on Recommender Systems (RecSys '23), September 18--22, 2023, Singapore, Singapore}
\acmDOI{10.1145/3604915.3610236}
\acmISBN{979-8-4007-0241-9/23/09}

\begin{document}

\title[Scaling Session-Based Transformer Recommendations]{Scaling Session-Based Transformer Recommendations using Optimized Negative Sampling and Loss Functions}

\author{Timo Wilm}
\authornote{Equal contribution}
\orcid{0009-0000-3380-7992}
\email{timo.wilm@otto.de}
\affiliation{%
  \institution{OTTO (GmbH \& Co KG)}
  \city{Hamburg}
  \country{Germany}
}
\author{Philipp Normann}
\authornotemark[1]
\orcid{0009-0009-5796-2992}
\email{philipp.normann@otto.de}
\affiliation{%
  \institution{OTTO (GmbH \& Co KG)}
  \city{Hamburg}
  \country{Germany}
}
\author{Sophie Baumeister}
\orcid{0009-0002-2988-6653}
\email{sophie.baumeister@otto.de}
\affiliation{%
  \institution{OTTO (GmbH \& Co KG)}
  \city{Hamburg}
  \country{Germany}
}
\author{Paul-Vincent Kobow}
\orcid{0009-0008-2610-3709}
\email{paul-vincent.kobow@otto.de}
\affiliation{%
  \institution{OTTO (GmbH \& Co KG)}
  \city{Hamburg}
  \country{Germany}
}

\acmArticleType{Research}

\acmCodeLink{https://github.com/otto-de/recsys-dataset}
\acmDataLink{https://www.kaggle.com/datasets/otto/recsys-dataset}




\begin{abstract}
  This work introduces \textbf{TRON}, a scalable session-based \textbf{T}ransformer \textbf{R}ecommender using \textbf{O}ptimized \textbf{N}egative-sampling. Motivated by the scalability and performance limitations of prevailing models such as SASRec and GRU4Rec\textsuperscript{+}, TRON integrates top-k negative sampling and listwise loss functions to enhance its recommendation accuracy. Evaluations on relevant large-scale e-commerce datasets show that TRON improves upon the recommendation quality of current methods while maintaining training speeds similar to SASRec. A live A/B test yielded an 18.14\% increase in click-through rate over SASRec, highlighting the potential of TRON in practical settings. For further research, we provide access to our source code\footnote[1]{\url{https://github.com/otto-de/TRON}} and an anonymized dataset\footnote[2]{\url{https://github.com/otto-de/recsys-dataset}}.
\end{abstract}

\maketitle

\section{Introduction}

Personalized real-time recommendations are a critical feature for e-commerce platforms such as \textit{OTTO}. While recent advancements in deep learning models have offered promising results in session-based recommendations~\cite{kang_self-attentive_2018,de_souza_pereira_moreira_transformers4rec_2021, mei_lightweight_2022}, established systems like RNN based GRU4Rec\textsuperscript{+}~\cite{hidasi_recurrent_2018} and transformer-based SASRec~\cite{kang_self-attentive_2018} often struggle to maintain accuracy and scalability when dealing with large item sets. To address these limitations, we introduce TRON, a session-based transformer recommendation system built upon the original SASRec, that uses top-k negative sampling and a listwise loss function to enhance accuracy and training time significantly.

\section{Methods}

\subsection{Negative Sampling}

Session-based recommendation systems predict the next item a user will interact with based on their previous activities within a session. A session is a sequence of user-item interactions, represented as $s:=[i_{1}, i_{2},\ldots, i_{T-1}, i_{T}]$ where $T$ is the session length. The items a user has interacted with within a session are considered positive samples, denoted as $\mathcal{P}(s):=\bigcup\limits_{k=1}^{T} \lbrace i_k \rbrace$. In contrast, items that the user has not interacted with are called negative samples, represented as $\mathcal{N}_s:= \mathcal{I} \setminus \mathcal{P}(s)$, where $\mathcal{I}$ is the total set of available items.

Training a model to perform a next-item prediction across $\mathcal{I}$ is often unfeasible due to the large size of $\mathcal{I}$ in real-world scenarios. Consequently, a common approach is to train the model to distinguish between positive and negative samples, which can be achieved through negative sampling~\cite{rendle_bpr_2009}. A major challenge in negative sampling is efficiency. Sampling directly from $\mathcal{N}_s$ can be computationally expensive, as it requires the exclusion of items present in $\mathcal{P}(s)$. This issue becomes critical when increasing the number of negative samples, leading to extended training times~\cite{kang_self-attentive_2018}.

An often utilized solution is to sample negatives according to a uniform distribution $\mathcal{U_I}$ across the entire set of items $\mathcal{I}$~\cite{hidasi_recurrent_2018}. This strategy proves to be effective for large item sets, where the probability of mistakenly sampling a positive item as a negative is relatively small. Another strategy is to sample negatives from the empirical frequency $\mathcal{F_I}$ of item interactions across all users. One method to efficiently and effectively sample negatives from $\mathcal{F_I}$, is in-batch negative sampling~\cite{hidasi_session-based_2016}. This method involves sampling negatives from the batch currently being processed. This is possible in GRU4Rec because, due to the way a batch is constructed, at each time step $t$, no other item from the session $s$ except $i_t$ exists in the batch. For transformer-based models, such as SASRec, we have developed an efficient solution to employ in-batch negative sampling by excluding samples from $s$ for batches that include all events of $s$.

In practice, the combination of negative sampling from both $\mathcal{U_I}$ and $\mathcal{F_I}$ often results in enhanced model accuracy~\cite{hidasi_recurrent_2018}. This can be achieved by sampling $k$ negatives from $\mathcal{U_I}$ and $m$ negatives from $\mathcal{F_I}$. Consider a batch $\mathcal{B}:= [S_1, S_2,\ldots, S_b]$ that consists of $b$ user sessions. At each time step $t$ in each user session $s$, we sample $\mathcal{UN}_s^t:=[U_1, U_2,\ldots, U_k]$ and $\mathcal{FN}_s^t:=[F_1, F_2,\ldots, F_m]$, where each $U_i$ is a sample from $\mathcal{U_I}$ and each $F_j$ is a sample from $\mathcal{F_I}$. These samples are then concatenated to form a $k+m$ dimensional random vector $\mathcal{N}_s^t:= concat[\mathcal{FN}_s^t, \mathcal{UN}_s^t]$. The negative samples for the entire batch are represented as $\mathcal{N}$.

This sampling process can be performed in different ways: elementwise, sessionwise, or batchwise, or a combination of these methods. For the elementwise approach, $\mathcal{N}$ is a tensor of shape $[b, T, k+m]$ because negatives are sampled at each time step for each session. In sessionwise sampling, all negatives for a session are sampled at once, resulting in a tensor of shape $[b,1,k+m]$. With batchwise sampling, all negatives for a batch are sampled at once, leading to a tensor of shape $[1,1,k+m]$. These different sampling strategies have a significant impact on the speed of training. For instance, when a large number of negatives is used, the data transfers between the CPU and GPU can become a bottleneck, particularly with elementwise sampling. Employing sessionwise or batchwise sampling can mitigate this issue, allowing the use of more negative samples per time step while maintaining a training speed comparable to that of SASRec. TRON uses a combination of uniform batchwise and in-batch sessionwise negative sampling to maintain training speed while improving accuracy.

To further optimize the negative sampling process and enhance recommendation performance TRON utilizes a \textbf{top-k negative sampling} strategy, which is inspired by a participant's idea from \textit{OTTO's} RecSys competition on \textit{Kaggle}\footnote[1]{\url{https://kaggle.com/competitions/otto-recommender-system/discussion/384022}} and is similar to dynamic negative item sampling \cite{zhang_optimizing_2013}. This strategy focuses on updating the top-k negatives during training instead of updating the whole set of negative ratings.

Initially, we sample a set of negative items $\mathcal{N}_{s}^{t}$ and obtain scores $r_{s,j}^t$ for each item $j$ of session $s$ at time step $t$ in $\mathcal{N}_{s}^{t}$. Applying the top-k function to the scored items, we select the top-k negatives $\mathcal{KN}_{s}^{t} := topk([r_{s,1}^t, r_{s,2}^t, \ldots, r_{s,|\mathcal{N}_{s}^{t}|}^t])$. These top-k items are then used for updates in the backpropagation step, while the rest are discarded.

This strategy allows us to retain the benefits of a large negative sample set, which provides a broader context and helps in identifying harder negatives, while substantially reducing the computational load during backpropagation. By prioritizing the update of negatives that the model currently misranks as likely positives, we enhance the overall speed and accuracy of the recommender system.

\subsection{Loss Functions}
Finally, we evaluate pointwise, pairwise, and listwise ranking loss functions typically used in recommendation systems~\cite{chen_ranking_2009} to further enhance model accuracy. The pointwise loss function is binary cross-entropy (BCE)~\cite{kang_self-attentive_2018}. The pairwise loss function is Bayesian personalized ranking max (BPR-MAX)~\cite{hidasi_recurrent_2018}. TRON uses sampled softmax (SSM)~\cite{bengio_quick_2003,bengio_adaptive_2008}, a listwise loss function with several beneficial properties, such as alleviating popularity bias and maximizing ranking metrics~\cite{wu_effectiveness_2022}.

\section{Experimental Setup}

\label{subsec:datasets}

\begin{table}
  \centering
  \caption{Statistics of the datasets used in our experiments.}
  \setlength{\tabcolsep}{2.75pt}
  \begin{tabular}{ lrrrrrrr }
    \toprule
    \multirow{3}{*}{\textbf{Data}} & \multicolumn{2}{c}{\textbf{Train set}} &                            & \multicolumn{2}{c}{\textbf{Test set}} &                              & \multirow{3}{*}{\textbf{items}}           \\
    \cmidrule{2-3} \cmidrule{5-6}
                                   & \multicolumn{1}{c}{sessions}           & \multicolumn{1}{c}{events} &                                       & \multicolumn{1}{c}{sessions} & \multicolumn{1}{c}{events}      &  &      \\
    \midrule
    Diginetica                     & 187k                                   & 906k                       &                                       & 18k                          & 87k                             &  & 43k  \\
    Yoochoose                      & 7,9M                                   & 31,6M                      &                                       & 15k                          & 71k                             &  & 37k  \\
    OTTO                           & 12,9M                                  & 194,7M                     &                                       & 1,6M                         & 12,3M                           &  & 1,8M \\
    \bottomrule
  \end{tabular}
  \label{tab:datasets}
\end{table}

In our evaluation, we assess the performance of our proposed model TRON, which is built upon the SASRec architecture, using three benchmark datasets: Diginetica~\cite{diginetica_cikm_2016}, Yoochoose~\cite{ben-shimon_recsys_2015}, and OTTO~\cite{philipp_normann_otto_2023}. Each of these datasets presents increasing complexity regarding the number of events and the variety of item sets. We only use click events for our experiments, maintaining a minimum item support of five and a session length of at least two for all datasets~\cite{hidasi_recurrent_2018}. We use a temporal train/test split method, using the last day (Yoochoose dataset) or the last week of data (Diginetica and OTTO datasets) to form the test sets. The remaining data is used for training. Table~\ref{tab:datasets} provides an overview of the datasets used in our experiments.

Recall@20 and MRR@20 are used as offline metrics~\cite{hidasi_session-based_2016}. We perform extensive assessments encompassing all events and items within the test set to ensure rigorous and dependable evaluations. We prioritize such comprehensive evaluations over sampling-based approaches because the latter have shown to be unreliable~\cite {krichene_sampled_2020}.

We use GRU4Rec\textsuperscript{+} and SASRec as benchmark models. GRU4Rec\textsuperscript{+} operates with a hidden size of 100. SASRec is configured with a hidden size of 200 across two layers. We introduce modifications to SASRec with two variant configurations: one with 512 uniform and 16 in-batch sessionwise negatives (SASRec M-Negs) and the other with 8192 uniform and 127 in-batch sessionwise negatives (SASRec L-Negs). Additionally, SASRec BPR-MAX adopts the BPR-MAX loss, while SASRec SSM leverages an SSM loss function, both utilizing the same negative sampling strategy as SASRec L-Negs. Our proposed models TRON L-Negs and XL-Negs are both based on the SASRec architecture and use an SSM loss function. TRON L-Negs is configured with 8192 batchwise uniform and 127 in-batch sessionwise negatives, whereas TRON XL-Negs operates with 16384 batchwise uniform negatives and 127 in-batch sessionwise negatives. Both TRON models use a top-k negative sampling strategy only updating based on the top 100 negative ratings. All models are trained with a batch size of 128 using an NVIDIA Tesla V100 GPU.

\section{Results}

\begin{table*}
  \centering
  \caption{Accuracy and training speed using various negative sampling strategies and loss functions. SASRec and TRON models were trained for 100 epochs on Diginetica, and 10 for both Yoochoose and OTTO, while GRU4Rec\textsuperscript{+} was trained for 10 epochs on Diginetica, 3 epochs on Yoochoose, and 1 epoch on OTTO. The best result for each dataset is highlighted in bold.}
  \setlength{\tabcolsep}{2.75pt}
  \begin{tabular}{ lrrrrrrrrrrr }
    \toprule
    \multirow{3}{*}{\textbf{Method}}       & \multicolumn{3}{c}{\textbf{Diginetica}} &                            & \multicolumn{3}{c}{\textbf{Yoochoose}} &  & \multicolumn{3}{c}{\textbf{OTTO}}                                                                                                                                                       \\
    \cmidrule{2-4} \cmidrule{6-8} \cmidrule{10-12}
                                           & \multicolumn{1}{c}{R@20}                & \multicolumn{1}{c}{MRR@20} & \multicolumn{1}{c}{Epochs/h}           &  & \multicolumn{1}{c}{R@20}          & \multicolumn{1}{c}{MRR@20} & \multicolumn{1}{c}{Epochs/h} &  & \multicolumn{1}{c}{R@20} & \multicolumn{1}{c}{MRR@20} & \multicolumn{1}{c}{Epochs/h} \\
    \midrule
    GRU4Rec\textsuperscript{+}             & 0.455                                   & 0.144                      & 15.126                                 &  & 0.725                             & \textbf{0.31}              & 0.478                        &  & 0.443                    & 0.205                      & 0.019                        \\
    SASRec                                 & 0.454                                   & 0.157                      & \textbf{94.533}                        &  & 0.573                             & 0.216                      & 2.573                        &  & 0.307                    & 0.180                      & \textbf{0.248}               \\
    \midrule
    SASRec M-Negs                          & 0.464                                   & 0.160                      & 93.581                                 &  & 0.607                             & 0.234                      & \textbf{2.603}               &  & 0.269                    & 0.142                      & 0.246                        \\
    SASRec L-Negs                          & 0.467                                   & 0.161                      & 48.247                                 &  & 0.571                             & 0.211                      & 1.245                        &  & 0.226                    & 0.114                      & 0.204                        \\
    \midrule
    SASRec BPR-Max                         & 0.526                                   & 0.175                      & 40.608                                 &  & 0.722                             & 0.297                      & 1.049                        &  & 0.377                    & 0.178                      & 0.194                        \\
    SASRec SSM                             & 0.516                                   & 0.169                      & 46.364                                 &  & 0.722                             & 0.305                      & 1.268                        &  & 0.432                    & 0.201                      & 0.209                        \\
    \midrule
    TRON L-Negs                            & 0.537                                   & 0.182                      & 81.389                                 &  & 0.730                             & 0.299                      & 2.117                        &  & 0.460                    & 0.212                      & 0.233                        \\
    TRON XL-Negs                           & \textbf{0.541}                          & \textbf{0.182}             & 68.408                                 &  & \textbf{0.732}                    & 0.302                      & 1.912                        &  & \textbf{0.472}           & \textbf{0.219}             & 0.227                        \\
    \midrule
    TRON XL vs. SASRec                     & 19.1\%                                  & 15.9\%                     & -27.6\%                                &  & 27.7\%                            & 39.8\%                     & -25.7\%                      &  & 53.7\%                   & 21.7\%                     & -8.5\%                       \\
    TRON XL vs. GRU4Rec\textsuperscript{+} & 18.9\%                                  & 26.4\%                     & 352.3\%                                &  & 0.97\%                            & -2.6\%                     & 299.8\%                      &  & 6.5\%                    & 6.8\%                      & 1094.7\%                     \\
    \bottomrule
  \end{tabular}
  \label{tab:results}
\end{table*}

The \textbf{offline evaluation} of our experiments compared to the benchmark models is presented in Table \ref{tab:results}. The GRU4Rec\textsuperscript{+} model outperforms SASRec across all datasets except MRR@20 on the Diginetica dataset. While previous studies on smaller datasets such as Diginetica indicated SASRec's superiority over GRU4Rec\textsuperscript{+}\cite{he_query-aware_2022, li_repetition_2023}, our findings on larger datasets and realistic training times do not support this claim. This discrepancy could also be attributed to our extensive evaluation method, which avoids weaknesses associated with sampling-based evaluations~\cite{krichene_sampled_2020} and does not solely rely on the last item of a session. SASRec M-Negs improves the accuracy of SASRec for the Diginetica and Yoochoose datasets but shows lower accuracy for the OTTO dataset while maintaining SASRec's original speed. SASRec L-Negs, on the other hand, exhibits slower training times across all datasets and only improves accuracy on Diginetica. This suggests that using additional negatives in a pointwise loss function such as BCE negatively impacts the model's accuracy. SASRec SSM shows promising results, outperforming GRU4Rec\textsuperscript{+} on the Diginetica dataset and demonstrating competitive accuracy for the other two datasets. Our proposed model TRON shows superior accuracy across all datasets except for MRR@20 on the Yoochoose dataset while demonstrating faster training times than SASRec SSM due to batchwise and top-k negative sampling. TRON demonstrates improved scalability as the dataset grows larger, as evidenced by the decreasing relative slowdown compared to SASRec. On the OTTO dataset, TRON shows an accuracy increase of more than 6.5\% in both Recall@20 and MRR@20, as well as a training speedup of 1090\% compared to GRU4Rec\textsuperscript{+}. Despite handling more negatives, TRON maintains 92\% of SASRec's training speed.

\begin{figure}
  \centering
  \includegraphics{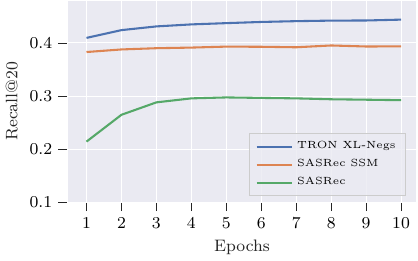}
  \caption{Offline evaluation results on our private OTTO dataset used for the online A/B test of our three groups.}
  \label{fig:recall}
  \Description[A line chart]{A line chart showing the offline evaluation results. The x-axis shows the number of training epochs, and the y-axis shows the Recall@20. The three lines represent the three groups: TRON XL-Negs, SASRec SSM, and SASRec. The Recall@20 of TRON XL-Negs starts at 0.409 and increases to 0.444 after 10 epochs. The Recall@20 of SASRec SSM starts at 0.383 and increases to 0.393 after 10 epochs. The Recall@20 of SASRec starts at 0.21 and increases to 0.297 after 5 epochs and then decreases to 0.292 after 10 epochs}
\end{figure}

\begin{figure}
  \centering
  \includegraphics{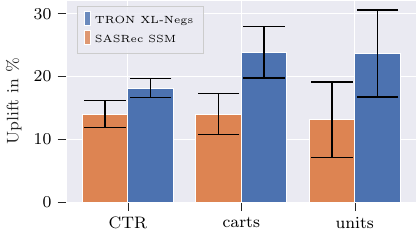}
  \caption{Online results of our A/B test relative to the SASRec baseline. The error bars indicate the 95\% confidence interval.}
  \label{fig:onex}
  \Description[A bar chart]{A bar chart showing the online results of TRON XL-Negs and SASRec SSM relative to SASRec. Three metrics are shown: click-through rate, carts, and units. In CTR TRON XL-Negs outperforms SASRec by 18.14\% (95\% confidence interval: 16.63\% to 19.67\%) and SASRec SSM outperforms SASRec by 14,17\% (95\% confidence interval: 12.68\% to 15.81\%). In carts TRON XL-Negs outperforms SASRec by 23.85\% (95\% confidence interval: 19.36\% to 27.53\%) and SASRec SSM outperforms SASRec by 14.04\% (95\% confidence interval: 11.27\% to 17.75\%). In units TRON XL-Negs outperforms SASRec by 23.67\% (95\% confidence interval: 16.34\% to 30.17\%) and SASRec SSM outperforms SASRec by 13.13\% (95\% confidence interval: 7.38\% to 19.43\%)}
\end{figure}

In the \textbf{online experiment}, we trained TRON XL-Negs, SASRec SSM, and SASRec on a private OTTO dataset from the two most recent weeks using the same preprocessing as described in Section \ref{subsec:datasets}. The Recall@20 for each epoch and model on the test set can be seen in Figure \ref{fig:recall}. The live improvement of TRON XL-Negs and SASRec SSM relative to SASRec measured from May 9 to May 17 2023 is shown in Figure \ref{fig:onex}. The results validate the effectiveness of TRON in a real-world e-commerce setting, showing an increase of 18.14\% in click-through rate, 23.85\% increase in add-to carts and 23.67\% uplift in units compared to SASRec.

\section{Conclusion}
Our proposed TRON model significantly improves the accuracy and training time of transformer-based recommendation systems on large e-commerce datasets. This enhancement is achieved through the strategic optimization of negative sampling methods, utilization of listwise loss functions, and focusing on the most misranked negatives.

\section{Speaker Bio}

\textbf{Timo Wilm}, \textbf{Philipp Normann}, and \textbf{Sophie Baumeister} form a data science trio at \textit{OTTO}'s recommendation team. Wilm and Normann are Senior Data Scientists with over five years of experience in e-commerce, specializing in the design and integration of cutting-edge deep learning models. Baumeister, a Junior Data Scientist, has been with the team for over one year. Together with their team, they are responsible for the development and maintenance of \textit{OTTO}'s recommendation systems, which are used by millions of customers every day.

\balance % Balance columns on last page

\bibliographystyle{ACM-Reference-Format}
\bibliography{paper}

\end{document}